\documentstyle[12pt]{article}
\begin{document}
\newcommand{\xy}{XY model}
\newcommand{\JW}{Jordan-Wigner}
\newcommand{\DM}{Dzyaloshinskii-Moriya}
\newcommand{\td}{thermodynamical}
\newcommand{\cm}{commutation rules}
\newcommand{\cf}{correlation function}
\newcommand{\op}{operator}
\newcommand{\sm}{statistical mechanics}
\newcommand{\Ham}{the Hamiltonian}
\newcommand{\tf}{transformation}
\newcommand{\cc}{calculation}
\newcommand{\sod}{$s=\frac{1}{2}$}
\newcommand{\od}{\mbox{$\frac{1}{2}$}}
\newcommand{\prim}{\prime}
\begin{center}
\large
Academy of Scinces of Ukraine\\
Institute for Condensed Matter Physics
\end{center}
\vspace{2 cm}
\large
\hspace*{9 cm}
Preprint\\
\hspace*{9 cm}
ICMP-93-22E\\
\vspace{2 cm}
\begin{center}
\Large
O.Derzhko, A.Moina\\
\vspace*{1 cm}
\LARGE
Statistical mechanics of
one-dimensional \sod\ anisotropic
       XY model in transverse field with \DM\ interaction\\
\vspace{6 cm}

\large
L'viv-1994
\end{center}

\clearpage
\noindent
O.Derzhko, A.Moina\\
\\
Statistical mechanics of one-dimensional \sod\ anisotropic
XY model in transverse field with \DM\ interaction\\
\\
For 1D \sod\ anisotropic \xy\ in transverse field with Dzyaloshinskii-Moriya
interaction using
\JW\ \tf\ the \td\ functions, static spin \cf s, transverse dynamical spin
\cf\ and connected with it transverse dynamical susceptibility have been
obtained. It has been shown that Dzyaloshinskii-Moriya interaction essentially
influences
the calculated quantities.\\
\\
\\
\\
\copyright 1994 Institute for Condensed Matter Physics
\normalsize
\clearpage
\section{Introduction}
	In 1961 E.Lieb,  T.Schultz and D.Mattis in ref. \cite{LSM} pointed out
one type of exactly solvable models of \sm\ that is so called 1D \sod\ \xy s.
Rewriting \Ham\ of such chain
\begin{equation}
H=\sum_{j}\left[ (1+\gamma)s_j^xs_{j+1}^x+(1-\gamma)s_j^ys_{j+1}^y\right],\;\;
-1\leq\gamma\leq 1,
\end{equation}
\begin{equation}
\left[ s_j^{\alpha},\:s_m^{\beta}\right]=\imath\delta_{jm}s_m^{\gamma},\;\;
\alpha,\beta,\gamma =x,y,z+ cyclic\; permutations
\end{equation}
with the help of the raising and lowering \op s
$s_j^{\pm}\equiv s_j^{x}\pm\imath s_j^{y}$ in the form
\begin{equation}
H=\od\sum_{j}\left[ (\gamma
s_j^{+}s_{j+1}^{+}+s_j^{+}s_{j+1}^{-})+h.c.\right],\label{quadr}
\end{equation}
they noted that the difficulty of diagonalization of the obtained quadratic
in operators $s^{+},s^{-}$ form (3) is connected with the commutation
rules that these operators satisfy, namely, $\left[s_j^{-},\:s_m^{+}\right]=
\delta_{jm}\left( 1-2s_m^{+}s_m^{-}\right)$. Really, they are similar to
Fermi-type \cm\ for \op s at the same site and to Bose-type \cm\ for operators
attached to different sites
\begin{eqnarray}
\left\{ s_j^{-},\:s_j^{+}\right\}=1,\;\;  (s_j^{+})^2=
(s_j^{-})^2=0;\nonumber\\
\nonumber\\
\left[s_j^{-},\:s_m^{+}\right]=\left[s_j^{+},\:s_m^{+}\right]=
 \left[s_j^{-},\:s_m^{-}\right]=0,\;\;j\neq m.
\end{eqnarray}
That is why one should perform at first \JW\  \tf\ (see, besides ref.
\cite{LSM}, also refs.[2-4])
\begin{eqnarray}
c_1=s_1^{-},\;\;c_j=s_j^{-}P_{j-1}=P_{j-1}s_j^{-},\;\; j=2,
\ldots,N,\nonumber\\
c_1^{+}=s_1^{+},\;\;c_j^{+}=s_j^{+}P_{j-1}=P_{j-1}s_j^{+},\;\;
j=2, \ldots ,N,
\end{eqnarray}
where \JW\ factor is denoted by $P_j\equiv\prod_{n=1}^j(-2s_n^z)$. The
introduced operators really obey Fermi \cm . From (5) it follows that
\begin{equation}
c_j^{+}c_j=s_j^{+}P_{j-1}^2s_j^{-}=s_j^{+}s_j^{-},\;\;
c_jc_j^{+}=s_j^{-}s_j^{+},\;\;c_j^{+}c_j^{+}=s_j^{+}s_j^{+},\;\;
c_jc_j=s_j^{-}s_j^{-},
\end{equation}
since $P_j^2=\prod_{n=1}^j(-2s_n^z)^2=\prod_{n=1}^j4(s_n^z)^2=1$, and the
\cm\  at the same site remain of Fermi-type. Consider then a product of
$c$-operators at different sites
\begin{equation}
c_n^{+}c_m=s_n^{+}\prod_{p=1}^{n-1}(-2s_p^{z})
\prod_{j=1}^{m-1}(-2s_j^{z})s_m^{-}=
s_n^{+}\prod_{j=n}^{m-1}(-2s_j^{z})s_m^{-},
\end{equation}
putting here for definiteness $n<m$. Since $s_j^{\pm}(-2s_j^z)=$
$\pm s_j^{\pm}$ and $ (-2s_j^z)s_j^{\pm}=\\ \mp s_j^{\pm}$, and consequently
\begin{equation}
c_mc_n^{+}=s_m^{-}\prod_{j=n}^{m-1}(-2s_j^{z})s_n^{+}=
-s_n^{+}\prod_{j=n}^{m-1}(-2s_j^{z})s_m^{-},
\end{equation}
one gets $c_n^{+}c_m=-c_mc_n^{+}$. Similarly one finds that
$c_n^{+}c_m^{+}=-c_m^{+}c_n^{+}$, $c_n c_m=\\=-c_m c_n$. Thus  the introduced
in (5) operators are Fermi-type operators
\begin{equation}
\left \{ c_j, \, c_l^{+} \right\}= \delta_{jl}, \;
\left\{ c_j^{+}, \,c_l^{+} \right\} = \left\{ c_j, \,c_l \right\} =0.
\end{equation}
Since $ P_j^2=1, P_j=\exp(\pm\imath\pi\!\sum_{n=1}^js_n^{+}s_n^{-})
$ (be\-cause $  \exp\left[\pm\imath\pi\sum_{n=1}^j(\od +s_n^z)\right]=\\
=\prod_{n=1}^j(-2s_n^z)$),
$s_j^{+}s_j^{-}=c_j^{+}c_j,$ it is easy to write the inverse to (5) \tf\
\begin{displaymath}
s_1^{-}=c_1,\;\;\;s_j^{-}=c_j\exp(\pm\imath\pi\sum_{n=1}^{j-1} c_n^{+}c_n) =
\exp(\pm\imath\pi\sum_{n=1}^{j-1} c_n^{+}c_n)c_j,\;\;\;
j=2, \ldots,N,
\end{displaymath}
\begin{equation}
s_1^{+}=c_1^{+},\;\;s_j^{+}=c_j^{+}\exp(\pm\imath\pi\sum_{n=1}^{j-1}
c_n^{+}c_n) =
\exp(\pm\imath\pi\sum_{n=1}^{j-1} c_n^{+}c_n)c_j^{+},\;\;\;
j=2, \ldots,N.
\end{equation}

	Returning to \Ham\ (3) one notes that the products of two
Pauli \op s at neighbouring sites transform into such products of Fermi \op s
\begin{eqnarray}
c_j^{+}c_{j+1}^{+}=s_j^{+}(-2s_j^{z})s_{j+1}^{+}=
s_j^{+}s_{j+1}^{+},\nonumber\\
c_j^{+}c_{j+1}    =s_j^{+}(-2s_j^{z})s_{j+1}^{-}=
s_j^{+}s_{j+1}^{-},\nonumber\\
c_j
c_{j+1}^{+}=s_j^{-}(-2s_j^{z})s_{j+1}^{+}=-s_j^{-}s_{j+1}^{+},\nonumber\\
c_j    c_{j+1}    =s_j^{-}(-2s_j^{z})s_{j+1}^{-}=-s_j^{-}s_{j+1}^{-}.
\end{eqnarray}
Usually bearing in mind the study of \td\ properties of the system that
requires the performance of \td\ limit $N\rightarrow\infty$, the periodic
boundary conditions are implied
\begin{equation}
s_{N+1}^{\alpha}\equiv s_1^{\alpha},\;\; \alpha=x,y,z.
\end{equation}
In connection with this in (3) there are products of the form
\begin{eqnarray}
s_{N}^{+}s_{N+1}^{+}=s_{N}^{+}s_{1}^{+}=s_{1}^{+}s_{N}^{+}=
c_{1}^{+}c_{N}^{+}\prod_{p=1}^{N-1}(-2s_p^{z})=c_{1}^{+}c_{N}^{+}P,\;\;
P\equiv P_N,\nonumber\\
s_{N}^{+}s_{N+1}^{-}= c_{1}    c_{N}^{+}P,\;\;
s_{N}^{-}s_{N+1}^{+}=-c_{1}^{+}c_{N}    P,\;\;
s_{N}^{-}s_{N+1}^{-}=-c_{1}    c_{N}    P.
\end{eqnarray}
Gathering the obtained terms one finds that for the ring
\begin{equation}
H=H^{-}+BP^{+}=H^{+}P^{+}+H^{-}P^{-}.
\end{equation}
Here
\begin{equation}
H^{\pm} \equiv \od\sum_{j=1}^N \left[
 (\gamma c^{+}_jc^{+}_{j+1}+c^{+}_jc_{j+1})+h.c. \right],
\end{equation}
the difference between $H^{+}$ and $H^{-}$ is only in the implied boundary
conditions:
for $H^{+}$ they are antiperiodic
\begin{equation}
c^{+}_j=-c^{+}_{j+N}, \; c_j=-c_{j+N},
\end{equation}
and for $H^{-}$ they are periodic
\begin{equation}
c^{+}_j=c^{+}_{j+N}, \; c_j=c_{j+N};
\end{equation}
 $B\equiv H^{+}-H^{-}=-\left[\left(\gamma c_N^{+}c_1^{+}+
c_N^{+}c_1\right)+h.c.\right] $ is the boundary term;
$ P^{\pm} \equiv (1\pm P)/2 $ are  orthogonal projectors ($P^{+}+P^{-}=1, \;
(P^{\pm})^2=P^{\pm},$ $ P^{\pm}P^{\mp}=0 $), besides this
$ \left [ H^{\pm},P \right ] = \left [ H^{\pm},P^{\pm} \right] = 0$.
For the open chain with free ends (then in the sum in (1) the range
of the summation index  is $j=1,\ldots,N-1$)  \Ham\ after fermionization
has the form
\begin{equation}
H =\od\sum_{j=1}^{N-1} \left[
 (\gamma c^{+}_jc^{+}_{j+1}+c^{+}_jc_{j+1})+h.c. \right].
\end{equation}

	Formulae (14), (15) or (18) realize the reformulation of the initial
Hamiltonian (1) in terms of fermions. They are the starting point for further
study of \sm\ of models like (1). Besides it appears [5,6] that for
calculation of free energy
\begin{equation}
f \equiv -\frac{1}{\beta} \lim_{N\rightarrow\infty} \left[ \frac{1}{N} \,
  Sp \exp(-\beta H) \right]
\end{equation}
or static spin \cf s
\begin{equation}
<s_{j_1}^{\alpha_1} \ldots s_{j_n}^{\alpha_n}>\equiv
\lim_{N\rightarrow\infty} \left\{Sp \left[\exp(-\beta H)
s_{j_1}^{\alpha_1} \ldots s_{j_n}^{\alpha_n} \right ]
/Sp \exp(-\beta H) \right\}
\end{equation}
the boundary term may be omitted and hence one has to consider a
system of free fermions. It is more difficult to calculate the dynamical
\cf s. Really,
\begin{eqnarray}
\lefteqn{s^z_j(t) \equiv \exp(\imath Ht)  s^z_j \exp (-\imath Ht)=}\nonumber\\
      &          & \!\!\!=\exp(\imath H^{+}t) s^z_j\exp (-\imath H^{+}t)P^{+}
                  +\exp(\imath H^{-}t) s^z_j \exp (-\imath
H^{-}t)P^{-}=\nonumber\\
      &          & \!\!\!=P^{+}\exp(\imath H^{+}t) s^z_j\exp (-\imath H^{+}t)
                  +P^{-}\exp(\imath H^{-}t) s^z_j \exp (-\imath H^{-}t)
\end{eqnarray}
(owing to the following relation that is valid for arbitrary function of
$H = \\ H^{+}P^{+}+H^{-}P^{-}$:
$f(H) \equiv
\sum_{n=0}^{\infty} \frac {f^{(n)}(0)}{n!}(H^{+}P^{+}+H^{-}P^{-})^n=
\sum_{n=0}^{\infty} \frac {f^{(n)}(0)}{n!}\times\\ \times
\!\left[(H^{+})^nP^{+}\!+\!(H^{-})^nP^{-} \right]\!
=\!f(H^{+})P^{+}+f(H^{-})P^{-}=
P^{+}f(H^{+})+P^{-}f(H^{-})$) in contrast to
\begin{eqnarray}
\lefteqn{s^{x,y}_j(t) =}\nonumber\\
  & &
    \!\!\!\!\!=\exp(\imath H^{+}t)\, s^{x,y}_j \exp (-\imath H^{-}t)P^{-}
     +\exp(\imath H^{-}t)\, s^{x,y}_j \exp (-\imath H^{+}t)P^{+}=\nonumber\\
  & &\!\!\!\!\! =P^{+}\!\exp(\imath H^{+}t)\, s^{x,y}_j\exp (-\imath H^{-}t)
     +P^{-}\!\!\exp(\imath H^{-}t)\, s^{x,y}_j \exp (-\imath H^{+}t).
\end{eqnarray}
In accordance with (21) the pair transverse \cf\ in the \td\ limit can be
written as
\begin{equation}
<s^z_j(t)s^z_{j+n}>= \frac{Sp \left [
\exp(-\beta H^{-})\exp(\imath H^{-}t)s^z_j\exp(-\imath H^{-}t)s^z_{j+n} \right
] }
  { Sp\,\exp \left (-\beta H^{-} \right ) }
\end{equation}
and hence may be calculated with $c$-cyclic Hamiltonian. Whereas the pair
longitudinal  \cf\ in accordance with (22) in the \td\ limit can be
written as
\begin{equation}
<s^x_j(t)s^x_{j+n}>= \frac{Sp \left [
         \exp(-\beta H^{-})\exp(\imath H^{-}t)s^x_j
         \exp(-\imath H^{-}t) O^{-}(t)s^x_{j+n} \right ] }
         { Sp\,\exp \left (-\beta H^{-} \right ) }
\end{equation}
where $O^{-}(t)\equiv \exp(\imath H^{-}t)\exp\left(
-\imath (H^{-}+B)t\right) $. The calculation with $c$-cyclic Hamiltonian that
neglects the boundary term $B$ yields the approximate result that, in
particular,
is incorrect in the limit of Ising model ($\gamma =1$) (see, for instance,
[7]). It is interesting to note that the calculation of the  four-spin \cf\
in the \td\ limit involves only $c$-cyclic Hamiltonian
\begin{eqnarray}
\lefteqn{<s^x_{j_1}(t)s^x_{j_2}(t)s^x_{j_3}s^x_{j_4}>=} \nonumber\\
&&=\frac{Sp \left [
         \exp(-\beta H^{-})\exp(\imath H^{-}t)s^x_{j_1}s^x_{j_2}
         \exp(-\imath H^{-}t) s^x_{j_3}s^x_{j_4} \right ] }
         { Sp\,\exp \left (-\beta H^{-} \right ) }.
\end{eqnarray}
Thus here as in the case (23) one comes to \cc\ of the dynamical \cf s of
the system of non-interacting fermions (see [8]). The \cc\ of the pair
longitudinal \cf , in spite of a great number of papers dealing with this
problem, remains an open point of \sm\ of 1D \sod\ \xy s. Among
other interesting and principal questions of the theory of 1D \sod\ \xy s one
may mention the investigation of nonequilibrium properties of such models
(see, for example, [9]) and the examination of the  properties of disordered
versions  of such models (see, for example, [10]).

	It is necessary to stress the  essential features of the present
consideration:
\begin{itemize}
  \item the dimension of space D=1;
  \item the value of spin \sod ;
  \item interactions occur only between neigh\-bouring spins
(other\-wise the Ha\-miltonian will contain  the terms  that are the products
of more than
two Fermi operators);
  \item only $x$ and $y$ components of spins interact  and the
field that may be included should be transverse (the interaction of $z$
components, for instance, leads
to the appearance in \Ham\  of the terms  that are the products of four Fermi
\op s).
\end{itemize}
In connection  with this it is easy  to point out the model that has more
general than in (1) form of interspin interaction, and that still allows the
described consideration. Really, considering the additional terms in \Ham\
that have form
\( \sum_j \left ( J^{xy}s^x_js^y_{j+1}+J^{yx}s^y_js^x_{j+1} \right ) \) one
notes that after fermionization they do not change the form  of \Ham\ (14),
(15)  or (18), and lead only to changes in the values  of constants. The
Hamiltonian of the generalized 1D \sod\ anizotropic \xy\ in transverse field
that as a matter of fact will be studied in the present paper is given by
\begin{equation}
H= \Omega \, \sum_js^z_j \, +
     \sum_j \left ( J^{xx}s^x_js^x_{j+1}+J^{xy}s^x_js^y_{j+1}+
                          J^{yx}s^y_js^x_{j+1}+J^{yy}s^y_js^y_{j+1} \right ).
\end{equation}

Before starting the examination of this model it is worthwhile to mention its
possible physical application [11]. For this purpose  let's perform the
\tf\ of rotation around axis $z$ over an angle $\alpha$
\begin{eqnarray}
  \tilde{s}_j^x=s_j^x \, \cos\alpha +s_j^y \, \sin\alpha, \; &
  \tilde{s}_j^y=-s_j^x \, \sin\alpha +s_j^y \, \cos\alpha, \; &
  \tilde{s}_j^z=s_j^z;\nonumber\\
   s_j^x=\tilde{s}_j^x \, \cos\alpha -\tilde{s}_j^y \, \sin\alpha, \;&
   s_j^y=\tilde{s}_j^x \, \sin\alpha +\tilde{s}_j^y \, \cos\alpha, \; &
   s_j^z=\tilde{s}_j^z.
\end{eqnarray}
Then rewritting at first new terms in sum in \Ham\ (26) in the form
\begin{equation}
 \frac{J^{xy}+J^{yx}}{2} \left (s^x_js^y_{j+1}+s^y_js^x_{j+1}\right )
+\frac{J^{xy}-J^{yx}}{2} \left (s^x_js^y_{j+1}-s^y_js^x_{j+1}\right ),
\end{equation}
taking into account that the terms
\( \left (s^x_js^y_{j+1}-s^y_js^x_{j+1}\right ) \) are invariant under
\tf\ (27) and
\begin{eqnarray}
J^{xx}s^x_js^x_{j+1}+
\frac{ J^{xy}+J^{yx} }{2}
\left( s^x_js^y_{j+1}+s^y_js^x_{j+1}\right)+
J^{yy}s^y_js^y_{j+1}=\nonumber\\
=\left( J^{xx} \cos^2\alpha + \frac{ J^{xy}+J^{yx} }{2} \sin{2\alpha}
+J^{yy} \sin^2\alpha\right)
\tilde{s}^x_j\tilde{s}^x_{j+1}+\nonumber\\+
\left( \frac{ J^{yy}-J^{xx} }{2}\sin{2\alpha}+
       \frac{ J^{xy}+J^{yx} }{2}\cos{2\alpha}\right)
\left(\tilde{s}^x_j\tilde{s}^y_{j+1}+\tilde{s}^y_j\tilde{s}^x_{j+1}\right)+\nonumber\\
+\left( J^{xx} \sin^2\alpha - \frac{ J^{xy}+J^{yx} }{2} \sin{2\alpha}
+J^{yy} \cos^2\alpha\right) \tilde{s}^y_j\tilde{s}^y_{j+1},
\end{eqnarray}
and choosing the parameter of \tf\ $\alpha$ from the condition
$ \left( J^{xy}+\right.$ \\ $\left. +J^{yx}\right) \cos{2\alpha}-$$
\left(J^{xx}-J^{yy}\right)$ $ \sin{2\alpha}=0$, one will have
\begin{equation}
H= \Omega \, \sum_j \tilde{s}^z_j \, +
     \sum_j\left [ J^{x}\tilde{s}^x_j\tilde{s}^x_{j+1}+
                          J^{y}\tilde{s}^y_j\tilde{s}^y_{j+1}
                          +D ( \tilde{s}^x_j\tilde{s}^y_{j+1}-
                               \tilde{s}^y_j\tilde{s}^x_{j+1}) \right ],
\end{equation}
where
\begin{eqnarray}
 J^{x}\equiv  J^{xx}\,\cos^2\alpha +\frac{ J^{xy}+J^{yx}}{2}\sin{2\alpha}
              + J^{yy}\,\sin^2\alpha,\nonumber\\
 J^{y}\equiv  J^{xx}\,\sin^2\alpha -\frac{ J^{xy}+J^{yx}}{2}\sin{2\alpha}
              + J^{yy}\,\cos^2\alpha,\nonumber\\
 D\equiv \frac{ J^{xy}-J^{yx}}{2},\;\;\;
 \tan{2\alpha}=\frac{ J^{xy}+J^{yx}}{ J^{xx}-J^{yy}}.
\end{eqnarray}
In the term that is proportional to $D$ one easily recognizes $z$ component of
the vector $[\vec{s}_j\times\vec{s}_{j+1}]$ that is  the so called
Dzyaloshinskii-Moriya
interaction. It was at first introduced phenomenologically by I.E.
Dzyaloshinskii
[12] and then derived by T.Moriya [13] by extending Anderson's theory of
superexchange
interactions [14] to include spin-orbital coupling (see, for example,
ref.[15]).
The model with relativistic \DM\ interaction together with ANNNI model are
widely
used in microscopic theory of crystals with incommensurate phase [16,17].
In the classical case \DM\ interaction may lead to the
appearance of the spiral spin structure. The possibility  of the appearance of
spiral structure in quantum case has been studied in ref.[11] where for this
purpose pair static spin \cf s have been estimated.

Except the mentioned paper [11] the problem of \sm\ of 1D \sod\ XY type
model with \Ham\ (26) or (30) as far as the authors know was not considered yet
\footnote{In ref.[18] on the base of the model with Hamiltonian (26) the
problem
about the validity of the Bose \cm\  approximation for spin \op s
has been examined.}.
At the present paper an attempt to fill up this gap by the generalization for
this case of the well-known scheme of consideration of 1D \sod\ \xy\ has been
made. In section 2 the \tf\ of  \Ham\ to the initial form for further
examination  of statistical properties is presented. In section 3 the
\td\ properties of the model are considered, and in section 4 it is shown how
to calculate the static spin \cf s in this model. The dynamics of transverse
spin correlations and the transverse dynamical susceptibility are studied in
section 5. The conclusions form section 6.
\section{ Transformation of \Ham\ }

In the spirit of above described approach the Hamiltionian  of the model (26)
at first should be rewritten with the help of the raising and lowering \op s
in the form that is similar to (3)
\begin{eqnarray}
\lefteqn{H= \Omega \, \sum_{j=1}^N \left (s^{+}_js^{-}_j -\frac{1}{2} \right
)+}\nonumber\\
&& + \sum_{j=1}^N \left (J^{++}s^{+}_js^{+}_{j+1}+J^{+-}s^{+}_js^{-}_{j+1}+
  J^{-+}s^{-}_js^{+}_{j+1}+J^{--}s^{-}_js^{-}_{j+1} \right ),\nonumber\\
&&  J^{++}\equiv\left [J^{xx}-J^{yy}-\imath (J^{xy}+J^{yx})\right ]/4
   =(J^{--})^{*},\nonumber\\
&& J^{+-}\equiv\left [J^{xx}+J^{yy}+\imath (J^{xy}-J^{yx})\right ]/4
   =(J^{-+})^{*};
\end{eqnarray}
here the periodic boundary conditions (12) are imposed. The Hamiltonian of
the model (32) after \JW\ \tf\ (5), (10)  will have the form that is similar
to (14),
(15)
\begin{eqnarray}
H=H^{+}P^{+}+H^{-}P^{-},\nonumber\\
H^{\pm}\equiv \Omega \, \sum_{j=1}^N \left (c^{+}_jc^{-}_j -\frac{1}{2} \right
) +
  \sum_{j=1}^N \left
(J^{++}c^{+}_jc^{+}_{j+1}+J^{+-}c^{+}_jc_{j+1}-\right.\nonumber\\
  \left. -J^{-+}c_jc^{+}_{j+1}-J^{--}c_jc_{j+1}\right );
\end{eqnarray}
besides $H^{-}$ is $c$-cyclic and $H^{+}$ is $c$-anticyclic quadratic forms in
Fermi
\op s. After Fourier \tf\
\begin{eqnarray}
c^{+}_{\kappa}=\frac{1}{\sqrt{N}}\sum_{j=1}^N e^{-\imath \kappa j}c^{+}_j,\;
c_{\kappa}=\frac{1}{\sqrt{N}}\sum_{j=1}^N e^{ \imath \kappa j}c_j,\;
 \nonumber\\
c^{+}_j=\frac{1}{\sqrt{N}}\sum_\kappa e^{\imath \kappa j}c^{+}_{\kappa},\;
c _j=\frac{1}{\sqrt{N}}\sum_\kappa e^{-\imath \kappa j}c_{\kappa},\;
\end{eqnarray}
with $\kappa=\kappa^{-}\equiv 2\pi n/N$ for $H^{-}$ and
$\kappa=\kappa^{+}\equiv 2\pi (n+1/2)/N$ for $H^{+}$, $n=-N/2, -N/2+1, \ldots,
N/2-1$
(for even N), $n=-(N-1)/2, -(N-1)/2+1, \ldots ,(N-1)/2$ (for odd N) $H^{\pm}$
can be rewritten in the form
\begin{eqnarray}
H^{\pm}=\sum_{\kappa} \left [ -\frac{\Omega}{2}+
        \epsilon_{\kappa}c^{+}_{\kappa}c_{\kappa}-
        \imath \sin{\kappa}\left (J^{++}c^{+}_{\kappa}c^{+}_{-\kappa}+
         J^{--}c_{\kappa}c_{-\kappa}\right )\right],\nonumber\\
\epsilon_{\kappa}\equiv \epsilon_{\kappa}^{(+)}+\epsilon_{\kappa}^{(-)},\;
\epsilon_{\kappa}^{(+)}\equiv\Omega+\frac{J^{xx}+J^{yy}}{2}\cos{\kappa},\;
\epsilon_{\kappa}^{(-)}\equiv \frac{J^{xy}-J^{yx}}{2}\sin{\kappa}\;
\end{eqnarray}
(here the following  relations
\begin{eqnarray}
\sum_{\kappa}e^{-\imath \kappa}\,c^{+}_{\kappa}c^{+}_{-\kappa}=
   -\imath \sum_{\kappa}\sin{\kappa} \,c^{+}_{\kappa}c^{+}_{-\kappa},\;\;
\sum_{\kappa}e^{ \imath \kappa}\,c_{\kappa}c_{-\kappa}=
   \imath \sum_{\kappa}\sin{\kappa}\, c_{\kappa}c_{-\kappa}
\end{eqnarray}
were used). The diagonalization of the quadratic forms is finished up by the
Bogolyubov \tf\
 \begin{eqnarray}
 \beta_{\kappa}=x_{\kappa}c_{\kappa}+y_{\kappa}c_{-\kappa}^{+},\; &
 \beta_{-\kappa}^{+}=y_{-\kappa}^{\ast}c_{\kappa}
           +x_{-\kappa}^{\ast}c_{-\kappa}^{+},\nonumber\\
  c_{\kappa}=\left (-x_{-\kappa}^{\ast}\beta_{\kappa}+
                     y_{\kappa}\beta_{-\kappa}^{+}\right )/\Delta_{\kappa},\;&
  c_{-\kappa}^{+}= \left (y_{-\kappa}^{\ast}\beta_{\kappa}-
                     x_{\kappa}\beta_{-\kappa}^{+}\right )/\Delta_{\kappa},
   \nonumber\\
   \Delta_{\kappa}\equiv
       y_{\kappa}y_{-\kappa}^{\ast}-x_{\kappa}x_{-\kappa}^{\ast}\neq 0.
\end{eqnarray}
$\beta$-operators remain of Fermi type if
\begin{equation}
  |x_{\kappa}|^2+|y_{\kappa}|^2=1,\;\;\;
  \frac{x_{\kappa}}{y_{\kappa}}+ \frac{x_{-\kappa}}{y_{-\kappa}}=0.
\end{equation}
The transformed Hamiltonian contains the \op\ terms  proportional only
to $\beta_{\kappa}^{+}\beta_{\kappa}$ if
\begin{equation}
\epsilon_{\kappa}^{(+)}
+\imath \sin{\kappa}\left (
 J^{++}\frac{x_{\kappa}}{y_{\kappa}}-
 J^{--}\frac{y_{-\kappa}}{x_{-\kappa}}\right )=0.
\end{equation}
The condition (39) and the second condition in (38) yield
\begin{eqnarray}
\frac{x_{\kappa}}{y_{\kappa}}=
  \imath \frac{\epsilon_{\kappa}^{(+)}\mp{\cal E}_{\kappa}}
              {2|J^{++}|\sin{\kappa}}
   \exp\left (-\imath argJ^{++}\right ),\nonumber\\
   {\cal
E}_{\kappa}\equiv\sqrt{(\epsilon_{\kappa}^{(+)})^2+4|J^{++}|^2\sin^2{\kappa}}.
\end{eqnarray}
Taking into account the first condition in (38) one finds that for lower sign
in (40)
\begin{eqnarray}
x_{\kappa}=2\imath |J^{++}|\sin{\kappa}\,\exp(-\imath argJ^{++})/
           \sqrt{2{\cal E}_{\kappa}({\cal
E}_{\kappa}-\epsilon_{\kappa}^{(+)})},\nonumber\\
y_{\kappa}=\sqrt{({\cal E}_{\kappa}-\epsilon_{\kappa}^{(+)})/2{\cal
E}_{\kappa}};
\end{eqnarray}
besides $\Delta_{\kappa}=1$, $E_\kappa=\epsilon_\kappa^{(-)}+{\cal E}_\kappa$.
For upper sign in (40)
\begin{eqnarray}
x_{\kappa}=-2\imath |J^{++}|\sin{\kappa}\,\exp(-\imath argJ^{++})/
           \sqrt{2{\cal E}_{\kappa}({\cal
E}_{\kappa}+\epsilon_{\kappa}^{(+)})},\nonumber\\
y_{\kappa}=\sqrt{({\cal E}_{\kappa}+\epsilon_{\kappa}^{(+)})/2{\cal
E}_{\kappa}};
\end{eqnarray}
besides $\Delta_{\kappa}=1$, $E_\kappa=\epsilon_\kappa^{(-)}-{\cal E}_\kappa$.
 Thus in a result of Bogolyubov  \tf\ (37) one gets
\begin{equation}
H^{\pm}=\sum_{\kappa}E_{\kappa}\left (
     \beta_{\kappa}^{+}\beta_{\kappa} -1/2 \right ),\;
      \left \{\beta_{\kappa},\:\beta_{\pi}^{+} \right \}=\delta_{\kappa \pi},\;
      \left \{\beta_{\kappa},\:\beta_{\pi} \right \}=
      \left \{\beta_{\kappa}^{+},\:\beta_{\pi}^{+} \right \}=0.
\end{equation}

	It is important to note that in contrast to anisotropic XY model
because of inequality $J^{xy}\neq J^{yx}$ one has $E_{\kappa}\neq E_{-\kappa}$.
This is connected with the absence of symmetry with respect to spatial
inversion.
Really, the Hamiltonian of the model (26)
$H(\Omega,J^{xx},J^{xy},J^{yx},J^{yy})$
under the action of spatial inversion, that leads to change of indexes $j$ to
$-j$
or $N-j$, $\:j+1$ to $N-j-1$, transforms into
$H(\Omega,J^{xx},J^{yx},J^{xy},J^{yy})$.

	In fig. 1 it is shown how the presence of Dzyaloshinskii-Moriya interaction
influences
the dependence of $E_\kappa=\epsilon^{(-)}_\kappa+{\cal E}_\kappa$ on $\kappa$
in de Gennes model
($1: D=0,\;\;\Omega=0,\;\;$ $ 1^{\prim}: D=J^{xx},\;\;\Omega =0;\;\;\;\;$
$2: D=0,\;\;\Omega=J^{xx},\;\;$ $ 2^{\prim}: D=J^{xx},\;\;\Omega =J^{xx}$).
In fig. 2 the same is depicted for the case of isotropic XY model.

	It is worthwhile to note that the spectrum of elementary excitations
in the model under consideration as it follows from the expression for ground
state
energy (46) is  given by $|E_\kappa|$.

\section{Thermodynamics}
For investigation of \td\ properties of the model in question let's calculate
the free energy per site in the limit $N\rightarrow\infty$. In accordance with
refs.[5,6] for such calculation one can use $c$-cyclic Hamiltonian and thus
\begin{equation}
f=-\frac{1}{\beta}\lim_{N\rightarrow \infty} \left [
       \frac{1}{N}\,\ln{Sp\, \exp{(-\beta H^{-})}} \right ].
\end{equation}
The diagonalized  quadratic in Fermi \op s form $H^{-}$ involved in (44)
has the form (43), and owing to this one easily obtains the desired result
\begin{eqnarray}
\lefteqn{f = - \frac{1}{\beta}\lim_{N\rightarrow\infty} \left \{
   \frac{1}{N}\ln{Sp\exp\left [-\beta \sum_{\kappa}E_{\kappa}\left (
       \beta_{\kappa}^{+}\beta_{\kappa}-\frac{1}{2}
         \right ) \right] }\right \}=}\nonumber\\
&& \!\!\!\!\!\!=-\frac{1}{\beta}\lim_{N\rightarrow\infty} \left \{
   \frac{1}{N}\ln{Sp\,\prod_{\kappa}\exp\left [-\beta  E_{\kappa}\left (
       \beta_{\kappa}^{+}\beta_{\kappa}-\frac{1}{2} \right)\
         \right]}\right\}=\nonumber\\
&& \!\!\!\!\!\!= -\frac{1}{\beta}\lim_{N\rightarrow\infty} \left [\!
     \frac{1}{N}\ln{\left (\prod_{\kappa}2 \!\cosh{\frac{\beta E_{\kappa}}{2}}
         \right ) }\! \right ] \!=\!
   -\frac{1}{\beta}\!\lim_{N\rightarrow\infty} \left [\!
     \frac{1}{N}\sum_{\kappa}\ln{(2\cosh{\frac{\beta E_{\kappa}}{2}})}
         \right ]=\nonumber\\
&&\!\!\!\!=-\frac{1}{\beta}\frac{1}{2\pi}\int_{-\pi}^{\pi}{d\kappa}\,\ln{\left
(2\,\cosh{
                \frac{\beta E_{\kappa}}{2}}\right ) }.
\end{eqnarray}

Knowing the free energy (45) one finds the energy of the ground state
\begin{equation}
e=\lim_{\beta\rightarrow\infty}f=-\frac{1}{4\pi}\int_{-\pi}^{\pi}
          d\kappa\, |E_{\kappa}|,
\end{equation}
the entropy
\begin{equation}
s=\beta^2\frac{\partial f}{\partial \beta}=\frac{1}{2\pi}\int_{-\pi}^{\pi}
          d\kappa\,\ln{\left (2\cosh{\frac{\beta E_{\kappa}}{2}}\right )}-
         \frac{1}{2\pi}\int_{-\pi}^{\pi} d\kappa\,
           \frac{\beta E_{\kappa}}{2} \,\tanh{
           \frac{\beta E_{\kappa}}{2}} ,
\end{equation}
the specific heat
\begin{equation}
c=-\beta\frac{\partial s}{\partial \beta}=\frac{1}{2\pi}\int_{-\pi}^{\pi}
           d\kappa\,
          {\left (\frac{\beta E_{\kappa}}{2}\right ) }^2 \, {\left (\cosh
           {\frac{\beta E_{\kappa}}{2}}\right ) }^{-2} ,
\end{equation}
the transverse magnetization
\begin{equation}
<\frac{1}{N}\sum_{j=1}^{N}s_j^z>=
  \frac{\partial f}{\partial \Omega}=-\frac{1}{4\pi}\int_{-\pi}^{\pi}
            d\kappa\,
           \frac{\partial E_{\kappa}}{\partial\Omega}\,\tanh{
           \frac{\beta E_{\kappa}}{2}} ,
\end{equation}
the static transverse susceptibility
\begin{eqnarray}
\lefteqn{\chi_{zz}=
   \frac{\partial<\frac{1}{N} \sum_{j=1}^N s_j^z>}{\partial
\Omega}=}\nonumber\\
 &&\!\!\!\!\!\!\!\!\!\!\!
    =         -\frac{1}{4\pi}\int_{-\pi}^{\pi} d\kappa\,
           \frac{\partial^2 E_{\kappa}}{\partial\Omega^2}
            \tanh{\frac{\beta E_{\kappa}}{2}} -
           \frac{\beta}{8\pi}\int_{-\pi}^{\pi} d\kappa\,
           {\left(\frac{\partial E_{\kappa}}{\partial\Omega}\right) }^2
            {\left(\cosh{\frac{\beta E_{\kappa}}{2}}\right ) }^{-2} .
\end{eqnarray}

 In order  to illustrate the influence of the additional  interactions on \td\
properties let's present the results of numerical calculations of the specific
heat (48)
as a function of temperature (figs. 3,4
(1: $D=0,\;\;\Omega=0,\;\;$ $ 1^{\prim}: D=J^{xx},\;\;\Omega =0;\;\;\;\;$
2: $D=0,\;\;\Omega=J^{xx},\;\;$ $ 2^{\prim}: D=J^{xx},\;\;\Omega =J^{xx}$))
and of transverse magnetization (49) as a function of transverse field
(figs. 5,6
(1: $D=0,\;\;\beta=1/J^{xx},\;\;$ $1^{\prim}:
D=J^{xx},\;\;\beta=1/J^{xx};\;\;\;\;$
2: $D=0,\;\;\beta=1000/J^{xx},\;\;$ $ 2^{\prim}:
D=J^{xx},\;\;\beta=1000/J^{xx}$))
and of temperature (figs. 7,8
(1: $D=0,\;\;\Omega=0,\;\;$ $ 1^{\prim}: D=J^{xx},\;\;\Omega =0;\;\;\;\;$
2: $D=0,\;\;\Omega=J^{xx},\;\;$ $ 2^{\prim}: D=J^{xx},\;\;\Omega =J^{xx}$))
for de Gennes model and for isotropic \xy\ in the presence of
Dzyaloshinskii-Moriya interaction.
\section{Static spin \cf s}

For the investigation of the spin stucture in the model under consideration
let's
introduce  the static spin \cf s. Due to the possibility of their calculation
with the help  of $c$-cyclic Hamiltonian the initial formula for their
evaluation
can be rewritten in the form
\begin{equation}
<s_{j_1}^{\alpha_1} \ldots s_{j_n}^{\alpha_n}>=
 \lim_{N\rightarrow \infty}\left\{ Sp\left[ \exp{(-\beta H^{-})}
    s_{j_1}^{\alpha_1} \ldots s_{j_n}^{\alpha_n} \right]/
   Sp\,\exp{(-\beta H^{-})}\right\}.
\end{equation}
Let's introduce  then $\varphi$-\op s that owing to (37), (41) are  the
following linear combinations of $\beta$-\op s
\begin{eqnarray}
\varphi_j^{\pm}\equiv c_j^{+} \pm c_j = \sum_{\kappa}
   \left( \lambda_{j\kappa}^{\pm}\beta_{\kappa}^{+}\pm
          \mu_{j\kappa}^{\pm}\beta_{-\kappa}\right),\nonumber\\
\lambda_{j\kappa}^{\pm}\equiv\frac{1}{\sqrt{N}}e^{\imath\kappa j}
          (x_{\kappa}\pm y_{\kappa}),\;\;
\mu_{j\kappa}^{\pm}\equiv\frac{1}{\sqrt{N}}e^{\imath\kappa j}
          [x_{\kappa}\exp{(2\imath \arg{J^{++}})}\pm y_{\kappa}],
\end{eqnarray}
and in terms of which the spin \op s can be presented as
\begin{equation}
s_n^x=\frac{1}{2}\prod_{j=1}^{n-1}\left( \varphi_j^{+}\varphi_j^{-}\right)
                     \varphi_n^{+},\;\;
s_n^y=\frac{1}{2\imath}\prod_{j=1}^{n-1}\left(
\varphi_j^{+}\varphi_j^{-}\right)
                     \varphi_n^{-},\;\;
s_n^z=\frac{1}{2}\varphi_n^{-}\varphi_n^{+}.
\end{equation}
$\varphi$-\op s obey the following commutation relations
\begin{equation}
\left \{\varphi_i^{+},\;\varphi_j^{-}\right\}=0,\;\;
\left \{\varphi_i^{+},\;\varphi_j^{+}\right\}=2\delta_{ij},\;\;
\left \{\varphi_i^{-},\;\varphi_j^{-}\right\}=-2\delta_{ij},
\end{equation}
besides $(\varphi_j^{+}\varphi_j^{-})^2 =1$ and
\begin{equation}
\left[ \varphi_i^{\pm},\,\varphi_j^{+}\varphi_j^{-}\right]=
       2\delta_{ij}\varphi_i^{\mp},\;\;
\left[ \varphi_i^{+}\varphi_i^{-},\,\varphi_j^{+}\varphi_j^{-}\right]=0.
\end{equation}
That is why the calculation of static spin \cf s after substitution of (53)
into (51) and exploiting of (54), (55) reduces to application
of Wick-Bloch-de Dominicis theorem. The theorem states that the mean value of
the product of  even number of $\varphi$ \op s with \Ham\ $H^-$ (43) is equal
to the
sum of all possible full systems of contractions of this product; if the number
of $\varphi$ \op s in the product is odd  the mean value of the
product is equal to zero. The full system of contractions of the product of
even number of Fermi-type  \op s forms so called Pfaffian the square of which
is equal to the determinant of antisymmetric matrix costructed in a certain
way from elementary contractions [19,20].

Thus let's consider the calculation of elementary contractions. One has
\begin{equation}
<\varphi_j^{+}\varphi_{j+n}^{+}>=
  \sum_{\kappa_1,\kappa_2}\left(
  \lambda_{j\kappa_1}^{+}\mu_{j+n,\kappa_2}^{+}
      <\beta_{\kappa_1}^{+}\beta_{-\kappa_2}>+
  \mu_{j\kappa_1}^{+}\lambda_{j+n,\kappa_2}^{+}
      <\beta_{-\kappa_1}\beta_{\kappa_2}^{+}>\right)
\end{equation}
(here evident relations $<\beta_{\kappa_1}^{+}\beta_{\kappa_2}^{+}>=
<\beta_{-\kappa_1}\beta_{-\kappa_2}>=0$ were used). Since
\begin{eqnarray}
<\beta_{\kappa_1}^{+}\beta_{-\kappa_2}>=\delta_{\kappa_1,-\kappa_2}/
         (1+e^{\beta E_{\kappa_1}})=\delta_{\kappa_1,-\kappa_2}
         f_{\kappa_1},\nonumber\\
<\beta_{-\kappa_1}\beta_{\kappa_2}^{+}>=\delta_{-\kappa_1,\kappa_2}
         e^{\beta E_{\kappa_2}}/
         (1+e^{\beta E_{\kappa_2}})=\delta_{-\kappa_1,\kappa_2}
         e^{\beta E_{\kappa_2}}
         f_{\kappa_2},
\end{eqnarray}
where $f_\kappa\equiv 1/(1+e^{\beta E_\kappa})$
and in accordance with (52)
\begin{eqnarray}
\lambda_{j\kappa}^{+}\mu_{j+n,-\kappa}^{+}=\frac{1}{N}e^{-\imath\kappa n}
            \left( 1+S_\kappa\right),\;&
\mu_{j,-\kappa}^{+}\lambda_{j+n,\kappa}^{+}=\frac{1}{N}e^{\imath\kappa n}
            \left( 1+S_\kappa\right),\nonumber\\
            S_\kappa\equiv \frac{2|J^{++}|\sin{(arg\,J^{++})}\sin{\kappa}}
                    {{\cal E}_{\kappa}}&
\end{eqnarray}
one has
\begin{equation}
<\varphi_j^{+}\varphi_{j+n}>=
\frac{1}{N}\sum_\kappa e^{\imath\kappa n}
             \left( 1+S_\kappa\right) -
        \frac{2\imath}{N}\sum_{\kappa}\sin{(\kappa n)}(1+S_\kappa )f_\kappa.
\end{equation}
Similarly one finds that
\begin{equation}
<\varphi_j^{-}\varphi_{j+n}^{-}>
=-\frac{1}{N}\sum_\kappa e^{\imath\kappa n}
             \left( 1-S_\kappa\right) +
        \frac{2\imath}{N}\sum_{\kappa}\sin{(\kappa n)}(1-S_\kappa )f_\kappa
\end{equation}
and
\begin{eqnarray}
\!\!\!\!\!\!\!\lefteqn{<\varphi_j^{+}\varphi_{j+n}^{-}>=}\nonumber\\
&& \!\!\!=\frac{1}{N}\sum_\kappa e^{\imath\kappa n}\!
        \left( \frac{\epsilon_\kappa^{(+)}}{{\cal E}_\kappa}+\imath C_\kappa
\!\right)
      \!  -\frac{2}{N}\sum_\kappa \left[\cos{(\kappa n)}\frac
         {\epsilon_\kappa^{(+)}}{{\cal E}_\kappa}-\sin{(\kappa n)}
          C_\kappa\! \right] f_\kappa ,
\end{eqnarray}
\begin{eqnarray}
\lefteqn{<\varphi_j^{-}\varphi_{j+n}^{+}>= }\nonumber\\
&&     =\frac{1}{N}\sum_\kappa e^{\imath\kappa n}
        \left( \frac{-\epsilon_\kappa^{(+)}}{{\cal E}_\kappa}+\imath C_\kappa
\right)
        +\frac{2}{N}\sum_\kappa \left[\cos{(\kappa n)}\frac
         {\epsilon_\kappa^{(+)}}{{\cal E}_\kappa}+\sin{(\kappa n)}\,
          C_\kappa \right] f_\kappa  ,\nonumber\\
&&        C_\kappa\equiv \frac{2|J^{++}|\cos{(arg\,J^{++})}\sin{\kappa}}
                    {{\cal E}_{\kappa}}.
\end{eqnarray}
The essential simplifications in expressions for contractions (59)-(62)
take place in the case of model (30), that is when $J^{xy}=-J^{yx}=D$. Then
\begin{eqnarray}
<\varphi_j^{+}\varphi_{j+n}^{+}>=\delta_{n,0}
     +\frac{\imath}{N}\sum_{\kappa}
     \sin{(\kappa n)}\tanh{\frac{\beta E_\kappa}{2}}\equiv E(n),\nonumber\\
<\varphi_j^{-}\varphi_{j+n}^{-}>=-E(n),\nonumber\\
<\varphi_j^{+}\varphi_{j+n}^{-}>=
     \frac{1}{N}\sum_{\kappa}
     \cos{(\kappa n+\psi_\kappa)}\tanh{\frac{\beta E_\kappa}{2}}\equiv
G(n),\nonumber\\
<\varphi_j^{-}\varphi_{j+n}^{+}>=-G(-n),
\end{eqnarray}
where $\cos{\psi_\kappa}\equiv\epsilon_\kappa^{(+)}/{\cal E}_\kappa,\;\;\
\sin{\psi_\kappa}\equiv\ 2J^{++}\sin{\kappa}/{\cal E}_\kappa.$

Let's return to the evaluation of equal-time  spin \cf s and consider,
for instance, $<s_j^xs_{j+n}^x>$. For this \cf\ with the utilization of
(53)-(55) one derives
\begin{equation}
<s_j^x s_{j+n}^x>=\frac{1}{4}<
    \varphi_{j}^{-}\varphi_{j+1}^{+}\varphi_{j+1}^{-}\varphi_{j+2}^{+}\ldots
    \varphi_{j+n-1}^{+}\varphi_{j+n-1}^{-}\varphi_{j+n}^{+}>,
\end{equation}
and after exploiting Wick-Bloch-de Dominicis theorem in r.h.s.  of (64)
for its  square one gets the following  expression
\begin{displaymath}
\left[4<s_j^x s_{j+n}^x>\right]^2=\\
\end{displaymath}
\begin{displaymath}
\vline
\begin{array}{@{}c@{\hspace{.3 mm}}c@{\hspace{.3 mm}}c@{\hspace{.3 mm}}
              c@{\hspace{.3 mm}}c@{\hspace{.3 mm}}
              c@{\hspace{.3 mm}}c@{\hspace{.3 mm}}c@{}}
0 &    -E(1)    & \ldots  &   -E(n-1)  & -G(-1) & -G(-2) & \ldots&  -G(-n)  \\
E(1)   &  0     &  \ldots &   -E(n-2)  & -G(0)  & -G(-1) & \ldots&  -G(-n+1)\\
E(2)   &  E(1)  & \ldots  &   -E(n-3)  & -G(1)  & -G(0)  & \ldots&  -G(-n+2)\\
\vdots &\vdots &\vdots &\vdots &\vdots &\vdots &\vdots &\vdots\\
E(n-1) &  E(n-2)& \ldots   &  0&  -G(n-2) &-G(n-3)   &  \ldots&  -G(-1)\\
G(-1)  &  G(0)  & \ldots   & G(n-2)  &0   & E(1)     &  \ldots&    E(n-1)\\
G(-2)  &  G(-1) & \ldots   & G(n-3)  &-E(1) & 0      &  \ldots&    E(n-2) \\
G(-3)  & G(-2)  &\ldots    & G(n-4)  &-E(2) & -E(1)  &  \ldots&    E(n-3) \\
\vdots &\vdots  &\vdots &\vdots &\vdots &\vdots &\vdots       &\vdots\\
G(-n)  & G(-n+1)&\ldots    & G(-1)  &-E(n-1) & -E(n-2)&\ldots&     0
\end{array}
\vline \, .
\end{displaymath}
\nopagebreak
\begin{equation}
\end{equation}
In a similar way for other pair spin correlators one obtains
\begin{displaymath}
\left[4\imath<s_j^x s_{j+n}^y>\right]^2=\\
\end{displaymath}
\begin{displaymath}
\vline
\begin{array}{@{\hspace{.03 mm}}c@{\hspace{.5 mm}}c@{\hspace{.1 mm}}
              c@{\hspace{.1 mm}}c@{\hspace{.5 mm}}c@{\hspace{.5 mm}}
              c@{\hspace{.1 mm}}c@{\hspace{.1 mm}}c@{\hspace{.03 mm}}}
0 & -E(1)       &\ldots & -E(n-1) & -E(-n) & -G(-1) &  \ldots&-G(-n+1)\\
E(1) & 0        &\ldots & -E(n-2)  & -E(n-1)  & -G(0)& \ldots&-G(-n+2)\\
E(2) & E(1)     &\ldots & -E(n-3)  & -E(n-2)  & -G(1)& \ldots&-G(-n+3)\\
\vdots &\vdots &\vdots &\vdots &\vdots &\vdots &\vdots &\vdots\\
E(n-1) & E(n-2) & \ldots   &0&  -E(1) &-G(n-2)       &  \ldots&-G(0)\\
E(n)  & E(n-1)  &\ldots& E(1)  &0       & -G(n-1)    &  \ldots&  -G(1)\\
G(-1)  & G(0)   &\ldots& G(n-2) &G(n-1) & 0          &   \ldots&  E(n-2) \\
G(-2)  & G(-1)  &\ldots&G(n-3)&G(n-2) & -E(1)    &     \ldots& E(n-3) \\
\vdots &\vdots &\vdots &\vdots &\vdots &\vdots &\vdots &\vdots\\
G(-n+1)  & G(-n+2) &\ldots&G(0) &G(1) & -E(n-2)&       \ldots& 0
\end{array}
\vline \, ,
\end{displaymath}
\nopagebreak
\begin{equation}
\end{equation}
\begin{equation}
<s_j^x s_{j+n}^z>=0;
\end{equation}
\begin{displaymath}
\left[4\imath<s_j^y s_{j+n}^x>\right]^2=\\
\end{displaymath}
\begin{displaymath}
\vline
\begin{array}{@{\hspace{.03 mm}}c@{\hspace{.6 mm}}c@{\hspace{.6 mm}}
              c@{\hspace{.6 mm}}c@{\hspace{.6 mm}}c@{\hspace{.6 mm}}
              c@{\hspace{.6 mm}}c@{\hspace{.6 mm}}c@{\hspace{.03 mm}}}
0 & -E(1)  & \ldots & -G(1) & -G(0) & -G(-1) &\ldots&-G(-n+1)\\
E(1) & 0 &   \ldots & -G(2)  & -G(1)  & -G(0) &\ldots&-G(-n+2)\\
E(2) & E(1) &\ldots & -G(3)  & -G(2)  & -G(1) &\ldots&-G(-n+3)\\
\vdots &\vdots &\vdots &\vdots &\vdots &\vdots &\vdots &\vdots \\
G(1) & G(2) &  \ldots   &0   & E(1) &E(2)  &     \ldots&   E(n)\\
G(0)  & G(1)  &\ldots& -E(1) & 0    & E(1) &     \ldots&   E(n-1)\\
G(-1)  & G(0) &\ldots& -E(2) &-E(1) & 0    &     \ldots&   E(n-2) \\
G(-2)  & G(-1)&\ldots& -E(3) &-E(2) & -E(1) &    \ldots&   E(n-3) \\
\vdots &\vdots &\vdots &\vdots &\vdots &\vdots &\vdots &\vdots\\
G(-n+1) & G(-n+2) &\ldots&-E(n) &-E(n-1) & -E(n-2)&\ldots& 0
\end{array}
\vline \, ,
\end{displaymath}
\nopagebreak
\begin{equation}
\end{equation}
\begin{displaymath}
\left[4<s_j^y s_{j+n}^y>\right]^2=\\
\end{displaymath}
\begin{displaymath}
\vline
\begin{array}{@{\hspace{.03 mm}}c@{\hspace{.5 mm}}c@{\hspace{.1 mm}}
              c@{\hspace{.1 mm}}c@{\hspace{.5 mm}}c@{\hspace{.5 mm}}
              c@{\hspace{.1 mm}}c@{\hspace{.1 mm}}c@{\hspace{.03 mm}}}
0 & -E(1) &\ldots & -E(n-1) & -G(1) & -G(0)    &\ldots&-G(-n+2)\\
E(1) & 0 & \ldots & -E(n-2)  & -G(2)  & -G(1)  &\ldots&-G(-n+3)\\
E(2) & E(1)& \ldots & -E(n-3)  & -G(3)  & -G(2)&\ldots&-G(-n+4)\\
\vdots &\vdots &\vdots &\vdots  &\vdots &\vdots &\vdots &\vdots\\
E(n-1) & E(n-2) &\ldots   &0&  -G(n) &-G(n-1) &\ldots&-G(1)\\
G(1)  & G(2)    &\ldots& G(n)  &0 & E(1)     & \ldots&  E(n-1)\\
G(0)  & G(1)    &\ldots& G(n-1) &-E(1) & 0   & \ldots&  E(n-2) \\
G(-1)  & G(0)   &\ldots&G(n-2)&-E(2) & -E(1) & \ldots&  E(n-3) \\
\vdots &\vdots &\vdots &\vdots &\vdots &\vdots &\vdots &\vdots\\
G(-n+2)  & G(-n+3) &\ldots&G(1) &-E(n-1) & -E(n-2)& \ldots & 0
\end{array}
\vline \, ,
\end{displaymath}
\nopagebreak
\begin{equation}
\end{equation}
\begin{equation}
<s_j^y s_{j+n}^z>=0;
\end{equation}
\begin{equation}
<s_j^z s_{j+n}^x>=0,
\end{equation}
\begin{equation}
<s_j^z s_{j+n}^y>=0,
\end{equation}
\begin{equation}
\left[4<s_j^z s_{j+n}^z>\right] ^2=
\left|
\begin{array}{cccc}
0 & E(n) & G(0) &  G(n)\\
-E(n)  &0& G(-n) &  G(0)\\
-G(0)& -G(-n)& 0 &  -E(n)\\
-G(n)& -G(0)&  E(n)&  0
\end{array}
\right|
{}.
\end{equation}

In figs. 9-12  the temperature dependences
(1: $D=0,\;\;\Omega=0,\;\;$ $ 1^{\prim}: D=J^{xx},\;\;\Omega =0;\;\;\;\;$
2: $D=0,\;\;\Omega=J^{xx},\;\;$ $ 2^{\prim}: D=J^{xx},\;\;\Omega =J^{xx}$)
and in figs. 13-16 the dependences on transverse field
(1: $D=0,\;\;\beta=1/J^{xx},\;\;$ $ 1^{\prim}:
D=J^{xx},\;\;\beta=1/J^{xx}\;\;\;\;\;$
$ 2: D=0,\;\;\beta=10/J^{xx},\;\;$ $ 2^{\prim}: D=J^{xx},\;\;\beta=10/J^{xx}$)
for some pair spin static \cf s are depicted. It is necessary
to underline the peculiarities caused by the presence of Dzyaloshinskii-Moriya
interaction.
First, only $<s_j^xs_{j+n}^z>,$ $ <s_j^ys_{j+n}^z>,$$ <s_j^zs_{j+n}^x>,$$
<s_j^zs_{j+n}^y>$ are equal to zero, but not $ <s_j^xs_{j+n}^y> $  and
$<s_j^ys_{j+n}^x>$. The last two correlators tend to zero when
$J^{xy}=J^{yx}=0$. In this case  $E(n)=0$ for $n\neq 0$ and hence (66) and (68)
may be
rewritten as determinants of matrices with only non-zero rectangle (but
not square) submatrices
on their diagonals; such determinants are equal to zero. Second, the
dependence of pair static \cf s
on $n$ is nonmonotonic (in accordance with ref.[11] this fact indicates
the appearance in the system of the incommensurate spiral spin structure).

\section{Dynamics of transverse spin correlations and dynamical
 transverse susceptibility}

Let's consider the dynamics of transverse spin correlations calculating  for
this purpose the transverse  time-dependent (dynamical) pair spin \cf\
$<s_j^z(t)s_{j+n}^z>$.
Due to the possibility of exploiting for its calculation $c$-cyclic
Hamiltonian $H^{-}$ (43) the evaluation of this \cf\  in accordance with (53)
and (52) reduces to estimation of dynamical \cf s density-density for the
system of non-interacting fermions
\begin{eqnarray*}
\lefteqn{4<s_j^z(t)s_{j+n}^z>=}\\
&&=\!\!\!\!\sum_{\kappa_1,\kappa_2,\kappa_3,\kappa_4}\!\!
  <\left[ \lambda^{+}_{j\kappa_1}\beta^{+}_{\kappa_1}(t)
         +\mu^{+}_{j\kappa_1}\beta_{-\kappa_1}(t)\right]
   \left[ \lambda^{-}_{j\kappa_2}\beta^{+}_{\kappa_2}(t)-
          \mu^{-}_{j\kappa_2}\beta_{-\kappa_2}(t)\right]\!\times\\
&& \times \left[ \lambda^{+}_{j+n,\kappa_3}\beta^{+}_{\kappa_3}
         +\mu^{+}_{j+n,\kappa_3}\beta_{-\kappa_3}\right]
   \left[ \lambda^{-}_{j+n,\kappa_4}\beta^{+}_{\kappa_4}-
          \mu^{-}_{j+n,\kappa_4}\beta_{-\kappa_4}\right] >=
\end{eqnarray*}
\begin{eqnarray}
 =  \sum_{\kappa_1,\kappa_2,\kappa_3,\kappa_4}
  \!\! \left[\! -\lambda^{+}_{j\kappa_1}\lambda^{-}_{j\kappa_2}
           \mu^{+}_{j+n,\kappa_3}\mu^{-}_{j+n,\kappa_4}\right.
          \!\!<\beta^{+}_{\kappa_1}\beta^{+}_{\kappa_2}
           \beta_{-\kappa_3}\beta_{-\kappa_4}>
           e^{\imath (E_{\kappa_1}+E_{\kappa_2})t}+\nonumber\\
         +\lambda^{+}_{j\kappa_1}\mu^{-}_{j\kappa_2}
           \lambda^{+}_{j+n,\kappa_3}\mu^{-}_{j+n,\kappa_4}
          <\beta^{+}_{\kappa_1}\beta_{-\kappa_2}
           \beta^{+}_{\kappa_3}\beta_{-\kappa_4}>
           e^{\imath (E_{\kappa_1}-E_{-\kappa_2})t}-\nonumber\\
           -\lambda^{+}_{j\kappa_1}\mu^{-}_{j\kappa_2}
           \mu^{+}_{j+n,\kappa_3}\lambda^{-}_{j+n,\kappa_4}
          <\beta^{+}_{\kappa_1}\beta_{-\kappa_2}
           \beta_{-\kappa_3}\beta^{+}_{\kappa_4}>
           e^{\imath (E_{\kappa_1}-E_{-\kappa_2})t}-\nonumber\\
          -\mu^{+}_{j\kappa_1}\lambda^{-}_{j\kappa_2}
           \lambda^{+}_{j+n,\kappa_3}\mu^{-}_{j+n,\kappa_4}
          <\beta_{-\kappa_1}\beta^{+}_{\kappa_2}
           \beta^{+}_{\kappa_3}\beta_{-\kappa_4}>
           e^{\imath (E_{\kappa_2}-E_{-\kappa_1})t}+\nonumber\\
          +\mu^{+}_{j\kappa_1}\lambda^{-}_{j\kappa_2}
           \mu^{+}_{j+n,\kappa_3}\lambda^{-}_{j+n,\kappa_4}
          <\beta_{-\kappa_1}\beta^{+}_{\kappa_2}
           \beta_{-\kappa_3}\beta^{+}_{\kappa_4}>
           e^{\imath (E_{\kappa_2}-E_{-\kappa_1})t}-\nonumber\\
     \left. -\mu^{+}_{j\kappa_1}\mu^{-}_{j\kappa_2}
           \lambda^{+}_{j+n,\kappa_3}\lambda^{-}_{j+n,\kappa_4}
          <\beta_{-\kappa_1}\beta_{-\kappa_2}
           \beta^{+}_{\kappa_3}\beta^{+}_{\kappa_4}>
           e^{-\imath (E_{-\kappa_1}+E_{-\kappa_2})t}\right] .
 \end{eqnarray}
In r.h.s. of (74) only non-zero averages of $\beta$-\op s are written down
and the following relations
\begin{equation}
 \beta^{+}_\kappa(t)=\beta^{+}_\kappa\exp{(\imath E_\kappa t)},\;\;
 \beta_\kappa(t)=\beta_\kappa\exp{(-\imath E_\kappa t)}
\end{equation}
were used. The  averages of $\beta$-\op s can be calculated using
Wick-Bloch-de Dominicis theorem, e.g.
\begin{eqnarray}
\lefteqn {       <\beta^{+}_{\kappa_1}\beta^{+}_{\kappa_2}
        \beta_{-\kappa_3}\beta_{-\kappa_4}>=}\nonumber\\
&&   =   - \frac{\delta_{\kappa_1,-\kappa_3}}{1+e^{\beta E_{\kappa_1}}}
        \frac{\delta_{\kappa_2,-\kappa_4}}{1+e^{\beta E_{\kappa_2}}}+
        \frac{\delta_{\kappa_1,-\kappa_4}}{1+e^{\beta E_{\kappa_1}}}
        \frac{\delta_{\kappa_2,-\kappa_3}}{1+e^{\beta
E_{\kappa_2}}}=\nonumber\\
&&        =-f_{\kappa_1}f_{\kappa_2}
         \delta_{\kappa_1,-\kappa_3}\delta_{\kappa_2,-\kappa_4}+
          f_{\kappa_1}f_{\kappa_2}
         \delta_{\kappa_1,-\kappa_4} \delta_{\kappa_2,-\kappa_3}
\end{eqnarray}
etc. After computation of these averages one finds that the coefficients near
the averages contain the following products
$\lambda^{+}_{j\kappa}\mu^{+}_{j+n,-\kappa},$
$ \lambda^{+}_{j+n,\kappa}\mu^{+}_{j,-\kappa},$
$ \lambda^{-}_{j\kappa}\mu^{-}_{j+n,-\kappa},$
$ \lambda^{-}_{j+n,\kappa}\mu^{-}_{j,-\kappa},$
$ \lambda^{+}_{j\kappa}\mu^{-}_{j+n,-\kappa},$
$ \lambda^{+}_{j+n,\kappa}\mu^{-}_{j,-\kappa},$
$ \lambda^{-}_{j\kappa}\mu^{+}_{j+n,-\kappa},$
$ \lambda^{-}_{j+n,\kappa}\mu^{+}_{j,-\kappa}.$
They can be found with the help of (52). For simplicity in what follows
their values will be used in the case when $J^{xy}=-J^{yx}=D$. Then
\begin{eqnarray}
      \lambda^{+}_{j\kappa}\mu^{+}_{j+n,-\kappa}=
      \frac{1}{N}e^{-\imath \kappa n}=
      \lambda^{-}_{j\kappa}\mu^{-}_{j+n,-\kappa} ,\nonumber\\
       \lambda^{+}_{j+n,\kappa}\mu^{+}_{j,-\kappa}=
      \frac{1}{N}e^{\imath \kappa n}=
      \lambda^{-}_{j+n,\kappa}\mu^{-}_{j,-\kappa} ,\nonumber\\
      \lambda^{+}_{j\kappa}\mu^{-}_{j+n,-\kappa}=
      \frac{1}{N}e^{-\imath (\kappa n+\psi_\kappa)},\;\;\;\;\;\;
      \lambda^{+}_{j+n,\kappa}\mu^{-}_{j,-\kappa}=
      \frac{1}{N}e^{\imath (\kappa n-\psi_\kappa)},\nonumber\\
      \lambda^{-}_{j\kappa}\mu^{+}_{j+n,-\kappa}=
      \frac{1}{N}e^{-\imath (\kappa n-\psi_\kappa)},\;\;\;\;\;\;
      \lambda^{-}_{j+n,\kappa}\mu^{+}_{j,-\kappa}=
      \frac{1}{N}e^{\imath (\kappa n+\psi_\kappa)}.
\end{eqnarray}
Gathering (74)-(77) together one derives the desired expression for transverse
time-dependent  \cf\ for the model (30)
\begin{eqnarray}
\lefteqn{4<s_j^z(t)s_{j+n}^z>=}\nonumber\\
&&=  \left[\frac{1}{N}\sum_\kappa
   \frac{\cosh{\left(-\imath E_\kappa t+\imath\kappa n
   +\frac{\beta E_\kappa}{2}\right)}}
        { \cosh{  \left(\frac{\beta E_\kappa}{2}\right)  } }\right]^2+
   \left[\frac{1}{N}\sum_\kappa
   \frac{\sinh{\left(\imath\psi_\kappa
   +\frac{\beta E_\kappa}{2}\right)}}
         {\cosh{\left(\frac{\beta E_\kappa}{2}\right)}}\right]^2-\nonumber\\
&&    -  \left[\frac{1}{N}\sum_\kappa
   \frac{\sinh{\left(-\imath E_\kappa t+\imath\kappa n +\imath\psi_\kappa
   +\frac{\beta E_\kappa}{2}\right)}}
         {\cosh{\left(\frac{\beta E_\kappa}{2}\right)}}\right]\times\nonumber\\
&&\times   \left[\frac{1}{N}\sum_\kappa
   \frac{\sinh{\left(-\imath E_\kappa t+\imath\kappa n -\imath\psi_\kappa
   +\frac{\beta E_\kappa}{2}\right)}}
         {\cosh{\left(\frac{\beta E_\kappa}{2}\right)}}\right].
\end{eqnarray}
Although $E_\kappa$ and $\cos{\psi_\kappa},$ $ \sin{\psi_\kappa}$ in (78) are
determined by formulae (43), (63) for the case $J^{xy}=-J^{yx}=D$ the obtained
result covers the case (26) as well. Remembering formulae (27) and (31) one
should simply use
$J^{xx}\cos^2\alpha $ $+\frac{
J^{xy}+J^{yx}}{2}\sin{2\alpha}+J^{yy}\,\sin^2\alpha$
instead of $J^{x}$,
$\;J^{xx}\sin^2\alpha -\frac{
J^{xy}+J^{yx}}{2}\sin{2\alpha}+J^{yy}\,\cos^2\alpha$
instead of $J^{y}$, and $\frac{ J^{xy}-J^{yx}}{2}$ instead of $D$ with
$\tan{\!2\alpha\!}=\left(J^{xy}+J^{yx}\right)/ \\/\left(J^{xx}-J^{yy}\right)$.
If one puts
$D=0$ in (78) it transforms into the  well-known result  obtained by
Th.Niemeijer
[21]. The depicted in figs. 17-20 dependence of the transverse dynamical
autocorrelation function (78) on time
($\beta=10/J^{xx};\;\;$ $1: D=0,\;\;\Omega=0,\;\;$ $ 1^{\prim}: D=J^{xx},\;\;
\Omega =0;\;\;\;\;$
2: $D=0,\;\;\Omega=J^{xx},\;\;$ $  2^{\prim}: D=J^{xx},\;\;\Omega =J^{xx}$)
shows substantial changes  caused by  Dzyaloshinskii-Moriya
interaction.

The dynamical susceptibility
\begin{equation}
\chi_{\alpha\beta}(\kappa,\omega)\equiv\sum_{n=1}^{N}
      e^{\imath\kappa n}\int_0^\infty dt
      e^{\imath (\omega+\imath\varepsilon)t}
      \frac{1}{\imath}<[s_j^{\alpha}(t),\:s_{j+n}^{\beta}]>
\end{equation}
is of great interest from the point of view of observable properties of the
system.
The obtained result (78) permits one to calculate  the transverse dynamical
susceptibility. Really, taking into account  the translation invariance
one gets
\begin{displaymath}
<[s_j^z(t),\:s_{j+n}^z]>=<s_j^z(t)s_{j+n}^z>-<s_j^z(-t)s_{j-n}^z>=
\end{displaymath}
\begin{eqnarray*}
=\imath\left[\frac{1}{N}\!\sum_{\kappa}\!\cos{(\kappa n-E_\kappa t)}\right]
      \left[\frac{1}{N}\!\sum_{\kappa}\!\sin{(\kappa n-E_\kappa t)}
                    \tanh{\frac{\beta E_\kappa}{2}}\right]-\nonumber\\
-\frac{\imath}{2}\left\{ \left[\!\frac{1}{N}\!\sum_{\kappa}\!
                    \cos{(\kappa n-E_\kappa t-\psi_\kappa)}\right.
                    \tanh{\frac{\beta E_\kappa}{2}}\right]\!
                    \left[\!\frac{1}{N}\!\sum_{\kappa}\!
                    \sin{(\kappa n-E_\kappa t+\psi_\kappa)}\right]+
\end{eqnarray*}
\begin{equation}
                   +\left[\frac{1}{N}\!\sum_{\kappa}\!
                    \cos{(\kappa n-E_\kappa t+\psi_\kappa)}
                    \tanh{\frac{\beta E_\kappa}{2}}\right]\!
        \left.      \left[\frac{1}{N}\!\sum_{\kappa}\!
                    \sin{(\kappa n-E_\kappa t-\psi_\kappa)}\right]\right\}.
\end{equation}
Using for summation over sites in (79) the lattice sum
$\frac{1}{N}\sum_{n=1}^N e^{\imath \kappa n}=\delta_{\kappa,0}$, evaluating the
integrals
over $t$ of the form
\begin{equation}
\int dt e^{\imath (\omega+F_\kappa+\imath\varepsilon)}=\frac{\imath}
                   {\omega+F_\kappa+\imath\varepsilon},
\end{equation}
bearing in mind the definition of functions $\cos{\psi_\kappa},$
$\sin{\psi_\kappa},$ and performing thermodynamical limit one ends up with
\begin{eqnarray}
\chi_{zz}(\kappa ,\omega)=\frac{1}{8\pi}\int_{-\pi}^{\pi}d\rho\left[
  \frac{1+\cos{(\psi_\rho+\psi_{\rho-\kappa})}}
       {E_{\rho-\kappa}-E_{\rho}-\omega-\imath\varepsilon}+\right.
  \frac{1-\cos{(\psi_\rho+\psi_{\rho-\kappa})}}
       {-E_{\kappa-\rho}-E_{\rho}-\omega-\imath\varepsilon}-\nonumber\\
 -\frac{1+\cos{(\psi_\rho+\psi_{\rho+\kappa})}}
       {-E_{\rho+\kappa}+E_{\rho}-\omega-\imath\varepsilon}-
 \left. \frac{1-\cos{(\psi_\rho+\psi_{\rho+\kappa})}}
       {E_{-\rho-\kappa}+E_{\rho}-\omega-\imath\varepsilon}\right]
   \tanh{\frac{\beta E_\rho}{2}}.
\end{eqnarray}
Using the relation
\begin{equation}
\frac{1}{F_\rho-\omega-\imath\varepsilon}={\cal P}\frac{1}{F_\rho-\omega}+
      \imath\pi\delta(F_\rho-\omega),
\end{equation}
for real and imaginary parts of transverse susceptibility one gets final
expressions
\begin{eqnarray}
Re\chi_{zz}(\kappa ,\omega)=\frac{1}{8\pi}{\cal P}\int_{-\pi}^{\pi}d\rho\left[
\! \!\frac{1+\cos{(\psi_\rho+\psi_{\rho-\kappa})}}
       {E_{\rho-\kappa}-E_{\rho}-\omega}+ \right.
  \frac{1-\cos{(\psi_\rho+\psi_{\rho-\kappa})}}
       {-E_{\kappa-\rho}-E_{\rho}-\omega}-\nonumber\\
 \!\!-\frac{1+\cos{(\psi_\rho+\psi_{\rho+\kappa})}}
       {-E_{\rho+\kappa}+E_{\rho}-\omega}-
\left.  \frac{1-\cos{(\psi_\rho+\psi_{\rho+\kappa})}}
       {E_{-\rho-\kappa}+E_{\rho}-\omega}\right]
   \tanh{\frac{\beta E_\rho}{2}},\;\;\;\;
\end{eqnarray}
\begin{eqnarray}
Im{\chi_{zz}(\kappa ,\omega)}=\frac{1}{8}\int_{-\pi}^{\pi}d\rho\left\{\right.
\left[1+\cos{(\psi_\rho+\psi_{\rho-\kappa})}\right]
 \delta (E_{\rho-\kappa}-E_{\rho}-\omega)+\nonumber\\
+\left[1-\cos{(\psi_\rho+\psi_{\rho-\kappa})}\right]
 \delta (-E_{\kappa-\rho}-E_{\rho}-\omega)-\nonumber\\
- \left[1+\cos{(\psi_\rho+\psi_{\rho+\kappa})}\right]
\delta (-E_{\rho+\kappa}+E_{\rho}-\omega)-\nonumber\\
-\left[ 1-\cos{(\psi_\rho+\psi_{\rho+\kappa})}\right]
 \left.\delta ( E_{-\rho-\kappa}+E_{\rho}-\omega)\right\}
\tanh{\frac{\beta E_\rho}{2}}.
\end{eqnarray}
These are the main results of the present paper.

It is useful to look at the particular case $\kappa=0$. In this case one has
\begin{eqnarray}
\chi_{zz}(0,\omega)=
\frac{1}{4\pi}\int_{-\pi}^{\pi}d\!\rho\sin^2{\psi_\rho}
\left[
\frac{1}{-E_{\rho}-E_{-\rho}-\omega-\imath\varepsilon}\right.-\nonumber\\
- \left.      \frac{1}{E_{\rho}+E_{-\rho}-\omega-\imath\varepsilon}\right]
  \tanh{\frac{\beta E_\rho}{2}}
\end{eqnarray}
so that, for instance,
\begin{equation}
Im\chi_{zz}(0,\omega)=-\frac{1}{4}\int_{-\pi}^{\pi}d\!\rho\sin^2{\psi_\rho}
  \delta(2{\cal E}_\rho-\omega)  \tanh{\frac{\beta E_\rho}{2}}.
\end{equation}
In the case of isotropic \xy\ with \DM\ interaction $\sin{\psi_\rho}=0$ and
$Im\chi_{zz}(0,\omega)=0$
as one should expect because in this case $\left[
\sum_{j=1}^{N}s_j^z,\:H\right]=0$.
In the case of de Gennes model with Dzyaloshinskii-Moriya interaction when
${\cal E}_\rho=\sqrt{\Omega^2+\Omega J\cos{\rho}+J^2/ 4}$ one can integrate
in (87) over $\rho$ using the relation
\begin{equation}
\delta(2{\cal E}_\rho-\omega)=\sum_{\rho_0}\frac{\delta(\rho-\rho_0)}
  {2\left| \frac{\partial{\cal E}_\rho}{\partial\rho}\right|  },
\end{equation}
where by $ \rho_0=\rho_0(\omega)$ the solutions of the equation
$2{\cal E}_{\rho_0}-\omega=0$ are denoted. This equation can be written in the
form
\begin{equation}
\cos{\rho_0}=\frac{\omega^2-J^2-4\Omega^2}{4\Omega J},
\end{equation}
and when $\omega$ satisfies inequalities
\begin{equation}
-1\leq\frac{\omega^2-J^2-4\Omega^2}{4\Omega J}\leq 1
\end{equation}
or for $\Omega,J>0$
\begin{equation}
|J-2\Omega|\leq\omega\leq J+2\Omega,
\end{equation}
equation (89) has two solutions in the interval of integration
$\rho_0\geq 0$ and $-\rho_0$ .
Besides $ \partial{\cal E}_\rho/\partial\rho=$
$-\Omega J\sin{\rho}/2{\cal E}_\rho$, $\sin{\psi_\rho}=
J\sin{\rho}/2{\cal E}_\rho$, so that in the case of de Gennes model
with Dzyaloshinskii-Moriya interaction  one gets the following final result
\begin{eqnarray}
\lefteqn{Im\chi_{zz}(0,\omega)=}\nonumber\\
&&=\left\{
 \begin{array}{lr}
   \frac{-J |\sin{\rho_0}|}{16{\cal E}_{\rho_0} \Omega}
      \left(\tanh{\frac{\beta E_{\rho_0} }{2}}
           +  \tanh{\frac{\beta E_{-\rho_0} }{2}}\right),\;
    $if$ \left|J-2\Omega\right|\leq\omega\leq J+2\Omega,&\\
    0,\;\; $otherwise$. &
   \end{array}
\right.
\end{eqnarray}
The presented in figs. 21,22 results of the numerical calculations of
frequency dependence of $Im\chi_{zz}(0,\omega)$ (86) for
de Gennes model with Dzyaloshinskii-Moriya interaction
($\beta=10/J^{xx};\;\;$  $ 1: D=0,\;\; $
$ 2: D=0.5J^{xx},\;\;\;\;$  3: $  D=J^{xx} $)
show that the presence of this interaction dramatically changes the frequency
dependence. This fact seems to be of great importance in connection with the
possible experimental prove of the presence in the system of
Dzyaloshinskii-Moriya interaction on the base of experimental measurements
of $Im\chi_{zz}(0,\omega)$.
\section{Conclusions}
Let's sum up the results of present study of statistical mechanics of 1D
\sod\ XY anisotropic ring in transverse field with
Dzyaloshinskii-Moriya interaction. This interaction keeps the model in
the class of 1D \sod\ \xy s because after fermionization of \Ham\ one is faced
with the quadratic in Fermi operators forms. However, after their
diagonalization
one finds that the spectrum $E_\kappa$ no longer is even function of $\kappa$.
This leads only to some technical complications in computations. The obtained
thermodynamical functions and static spin \cf s essentially depend on the value
of  Dzyaloshinskii-Moriya interaction. For instance, these interaction
decreases
the transverse magnetization at certain transverse field in de Gennes model
and in isotropic XY model (figs. 5,6). They lead
to appearance of non-zero spin correlators $<s_j^xs_{j+n}^y>$ and
$<s_j^ys_{j+n}^x>$ and to nonmonotonic dependence  of pair spin \cf s on $n$.
The evaluation of transverse dynamical \cf\ and the corresponding
susceptibility
shows that Dzyaloshinskii-Moriya interaction essentially influences on the
dynamics of transverse spin  correlations (figs. 17-20)
and drastically changes the dynamical susceptibility (figs. 21,22).

In addition it is necessary to note that if $J^{xy}=J^{yx}=0$ all obtained
results transform into the corresponding results for  anisotropic XY model.
Really, in this limit
$E_\kappa\rightarrow{\cal E}_\kappa\rightarrow
\sqrt{\epsilon_\kappa^{(+)2}+4(J^{++})^2\sin^2{\kappa}}$ and
$E_\kappa=E_{-\kappa}$. Due to this simplification the thermodynamical
functions
because of parity of integrands contain $2\int_0^\pi d\!\kappa(\ldots)$ instead
of $\int_{-\pi}^\pi d\!\kappa(\ldots)$. In contractions (59)-(63)
$<\varphi_j^{+}\varphi_{j+n}^{+}>\rightarrow \delta_{n,0}$,
$<\varphi_j^{-}\varphi_{j+n}^{-}>\rightarrow -\delta_{n,0}$,
$<\varphi_j^{+}\varphi_{j+n}^{-}>\rightarrow \frac{1}{\pi}\int_0^\pi
  d\!\kappa\cos{(\kappa n+\psi_\kappa)}\tanh{\frac{\beta E_\kappa}{2}}
  \equiv G(n),$
$<\varphi_j^{-}\varphi_{j+n}^{+}>\rightarrow -G(-n)$ so that
$4<s_j^xs_{j+n}^x>,$ $\;4<s_j^ys_{j+n}^y>,$ $\;4<s_j^zs_{j+n}^z>$
(but not their squares) can be rewritten  as $N\times N$ determinants and
$<s_j^xs_{j+n}^y>=<s_j^ys_{j+n}^x>=0.$ The transverse dynamical \cf\ transforms
into the obtained in ref.[21] expression.

The performed investigations follow earlier works [18,22-29] considering the
derivation of exact results
in statistical mechanics of 1D \sod\  systems  with Dzyaloshinskii-Moriya
interaction.

At last it should be mentioned that for a quite a lot of magnetic and
ferroelectric materials,  showing nearly 1D behavior above their ordering
temeperatures, a variety of experimental data are now available [30-40]
and thus theoretical investigations of statistical mechanics of 1D spin
models may be of great interest for clarifying whether
the properties of such simple spin models are capable to caricature the
measurements.
\\

The authors would like to thank J.Rossat-Mignod, F.P.Onufrieva and other
participants of the Ukrainian-French
Symposium "Condensed Matter: Science \&	 Industry" (Lviv, 20-27 February 1993)
for stimulating discussions. They would like to express the gratitude to
the participants  of the seminar of Quantum Statistics Department of ICMP
(11.05.1993) and to the participants of seminars of Laboratory for the
Theory of Model Spin Systems of this department for many helpful discussions.

\vspace{1 cm}
\begin{center}
{\LARGE Figures caption}\\
\end{center}
\vspace{1 cm}
Fig.1.\hspace{.5 cm}
$E_\kappa /J^{xx}=(\epsilon_\kappa^{(-)}+{\cal E}_\kappa)/J^{xx}$ vs.
$\kappa;\;\;
J^{yy}=0.$\\
\\
Fig.2.\hspace{.5 cm}
$E_\kappa /J^{xx}=(\epsilon_\kappa^{(-)}+{\cal E}_\kappa)/J^{xx}$ vs.
$\kappa;\;\;
J^{yy}=J^{xx}.$\\
\\
Fig.3.\hspace{.5 cm}
$c$ vs. $1/(\beta J^{xx});\;\;\;
J^{yy}=0.$\\
\\
Fig.4.\hspace{.5 cm}
$c$ vs. $1/(\beta J^{xx});\;\;\;
J^{yy}=J^{xx}.$\\
\\
Fig.5.\hspace{.5 cm}
$-<\frac{1}{N}\sum_{j=1}^{N}s_j^z>$ vs. $\Omega/J^{xx};\;\;\;
J^{yy}=0.$\\
\\
Fig.6.\hspace{.5 cm}
$-<\frac{1}{N}\sum_{j=1}^{N}s_j^z>$ vs. $\Omega/J^{xx};\;\;\;
J^{yy}=J^{xx}.$\\
\\
Fig.7.\hspace{.5 cm}
$-<\frac{1}{N}\sum_{j=1}^{N}s_j^z>$ vs. $1/(\beta J^{xx});\;\;\;
J^{yy}=0.$\\
\\
Fig.8.\hspace{.5 cm}
$-<\frac{1}{N}\sum_{j=1}^{N}s_j^z>$ vs. $1/(\beta J^{xx});\;\;\;
J^{yy}=J^{xx}.$\\
\\
Fig.9.\hspace{.5 cm}
$16<s_j^xs_{j+1}^x>^2$ vs. $1/(\beta J^{xx});\;\;\;
J^{yy}=0.$\\
\\
Fig.10.\hspace{.5 cm}
$16<s_j^xs_{j+1}^x>^2$ vs. $1/(\beta J^{xx});\;\;\;
J^{yy}=J^{xx}.$\\
\\
Fig.11.\hspace{.5 cm}
$16<s_j^xs_{j+1}^y>^2$ vs. $1/(\beta J^{xx});\;\;\;
J^{yy}=0.$\\
\\
Fig.12.\hspace{.5 cm}
$16<s_j^xs_{j+1}^y>^2$ vs. $1/(\beta J^{xx});\;\;\;
J^{yy}=J^{xx}.$\\
\\
Fig.13.\hspace{.5 cm}
$16<s_j^xs_{j+1}^x>^2$ vs. $\Omega/J^{xx};\;\;\;
J^{yy}=0.$\\
\\
Fig.14.\hspace{.5 cm}
$16<s_j^xs_{j+1}^x>^2$ vs. $\Omega/J^{xx};\;\;\;
J^{yy}=J^{xx}.$\\
\\
Fig.15.\hspace{.5 cm}
$16<s_j^xs_{j+1}^y>^2$ vs.$\Omega/J^{xx};\;\;\;
J^{yy}=0.$\\
\\
Fig.16.\hspace{.5 cm}
$16<s_j^xs_{j+1}^y>^2$ vs. $\Omega/J^{xx};\;\;\;
J^{yy}=J^{xx}.$\\
\\
Fig.17.\hspace{.5 cm}
$Re<s_j^z(t)s_j^z>$ vs. $J^{xx}t;\;\;\;
J^{yy}=0.$\\
\\
Fig.18.\hspace{.5 cm}
$Im<s_j^z(t)s_j^z>$ vs. $J^{xx}t;\;\;\;
J^{yy}=0.$\\
\\
Fig.19.\hspace{.5 cm}
$Re<s_j^z(t)s_j^z>$ vs. $J^{xx}t;\;\;\;
J^{yy}=J^{xx}.$\\
\\
Fig.20.\hspace{.5 cm}
$Im<s_j^z(t)s_j^z>$ vs. $J^{xx}t;\;\;\;
J^{yy}=J^{xx}.$  \\
\\
Fig.21.\hspace{.5 cm}
$-Im\chi_{zz}(0,\omega)$ vs.  $\omega/J^{xx};\;\;\;
\Omega/J^{xx}=0.25.$\\
\\
Fig.22.\hspace{.5 cm}
$-Im\chi_{zz}(0,\omega)$ vs.  $\omega/J^{xx};\;\;\;
\Omega/J^{xx}=0.5.$ \\
\\
Institute for Condensed Matter Physics\\
1 Svientsitskii Str\\
Lviv 290011\\
Ukraine\\
e-mail: icmp@sigma.icmp.lviv.ua\\

\begin{thebibliography}{99}
\bibitem{LSM} Lieb E., Schultz T., Mattis D.
Two soluble models of an antiferromagnetic chain//Annals of Physics.-1961.
  -{\bf 16},3.-P.407-466.
\bibitem{katsura} Katsura S. Statistical  mechanics of the anisotropic
linear Heisenberg model//Phys.Rev.-1962.-{\bf 127},5.-P.1508-1518.
\bibitem{Ches} Chesnut D.B., Suna A. Fermion behavior of one-dimensional
excitons// J.Chem.Phys.-1963.-{\bf 39},1.-P.146-149.
\bibitem{pikin} Pikin S.A., Tsukernik V.M. Thermodynamics of linear
spin chains in a transverse magnetic field//Zh.Eksp.Teor.Fiz.-1966.-
{\bf 50},5.-P.1377-1380.(in Russian)
\bibitem{Mazur1} Mazur P., Siskens Th.J. Time \cf s in the $a$-cyclic XY model.
I//Physica.-1973.-{\bf 69}.-P.259-272.
\bibitem{Mazur2} Siskens Th.J., Mazur P. Time-\cf s in the $a$-cyclic XY model.
II//Physica.-1974.-{\bf 71}.-P.560-578.
\bibitem{Gons} Goncalves L.L. Dynamics of the one-dimensional transverse
Ising model//J.Phys.A.-1980.-{\bf 13}.-P.223-236.
\bibitem{McCoy} McCoy B.M., Barouch E., Abraham D.B. Statistical mechanics
of the XY model. IV. Time-dependent  spin-correlation functions //
Phys.Rev.A.-1971.-{\bf 4},6.-P.2331-2342.
\bibitem{Dresd} Barouch E., Dresden M. Exact time-dependent analysis for the
one-dimensional XY model // Phys.Rev.Lett.-1969.-{\bf 23},3.-P.114-117.
\bibitem {Nishim} Nishimori H. One-dimensional XY model in Lorentzian random
field // Phys.Lett.A.-1984.-{\bf 100},5.-P.239-243.
\bibitem{Kont} Kontorovich V.M., Tsukernik V.M. Spiral structure in the
one-dimensional
spin chain // Zh.Eksp.Teor.Fiz.-1967.-{\bf 52},5.-P.1446-1453. (in Russian)
\bibitem{dzyal} Dzyaloshinsky I.E. A thermodynamical theory of
"weak" ferromagnetism in antiferromagnetic substances//Zh.Eksp.Teor.Fiz.-1957.-
{\bf 32},6.-P.1547-1562. (in Russian)
\bibitem{Moriya} Moriya T. Anisotropic superexchange interaction and weak
ferromagnetism//Phys.Rev.-1960.-{\bf 120},1.-P.91-98.
\bibitem{ And} Anderson P.W. New approach to the theory of superexchange
interactions//Phys.Rev.-1959.-{\bf 115},1.-P.2-13.
\bibitem{Mess} Messia A. Kvantovaya mehanika, tom 2.-Moskva: Nauka, 1979.-584
p.
(in Russian)
\bibitem{Izym1} Izyumov Yu.A., Ozerov R.P. Magnitnaya neitronografiya.-Moskva:
Nauka, 1966.-532 p. (in Russian)
\bibitem{Izym2} Izyumov Yu.A. Difraktsiya neitronov na dlinnoperiodicheskih
strukturah.-Moskva: Energoatomizdat, 1987.-200 p. (in Russian)
\bibitem{Derzh} Derzhko O.V., Levitskii R.R., Moina A.Ph. Approximation of
Bose commutation rules and elementary excitation spectrum in the spin systems
theory//Fiz.Kond.Sys.-1993.-1.-P.115-118.  (in Ukrainian)
\bibitem{Isi} Isihara A. Statistical physics.- New York-London: Academic
Press, 1971.
\bibitem{Bax} Baxter R.J. Exactly solved models in statistical mechanics.-
London New York Paris San Diego San Francisco Sao Paulo Sydney Tokyo Toronto:
Academic Press, 1982.
\bibitem{Niem} Niemeijer Th. Some exact calculations on a chain of spins \od\
//Physica.-1967.-{\bf 36},3.-P.377-419.
\bibitem {Kryv} Krivoruchko V.N. Magnon bound states in an anisotropic chain
of spins with the Dzyaloshinskii interaction//Fiz.Niz.Temp.-1986.-{\bf 12},8.-
P.872-876. (in Russian)
\bibitem{Zv1}Zvyagin A.A. The ground state structure of a spin chain with the
Dzyaloshinsky type interaction//Fiz.Niz.Temp.-1989.-{\bf 15},9.-P.977-979.
(in Russian)
\bibitem {Zv2} Zvyagin A.A.  Characteristics of two-sublattice spin chain
with coupling of Dzyaloshinsky type //Zh.Eksp.Teor.Fiz.-1990.-{\bf
98},4.-P.1396-1401.
(in Russian)
\bibitem{Zv3} Zvyagin A.A. The effect of impurities on planar structure of
spin chain with Dzyaloshinsky interaction//Fiz.Niz.Temp.-1991.-{\bf 17},1.-
P.125-127. (in Russian)
\bibitem{Zv4} Zvyagin A.A. The transverse structure of a spin chain with
Dzyaloshinskii-Moriya-type interaction //J.Phys.: Condens.Matter.-1991.-{\bf
3}.-
P.3865-3867.
\bibitem{de} Derzhko O.V., Smereka I.T., Moina A.Ph. Thermodynamical and
dynamical properties of XY-type 1D \sod -spin models and some applications
in condensed matter physics //Ukrainian-French
Symposium "Condensed Matter: Science \& Industry"
(Lviv, February, 20-27, 1993). Abstracts, information \&
participants.-Lviv,1993.-P.198.
\bibitem{derzh1} Derzhko O.V., Moina A.Ph. 1D \sod\ anisotropic \xy\ in
transverse field with Dzyaloshinskii-Moriya interaction //The Eighth
International
Meeting on Ferroelectricity, 8-13 August 1993. Program Summary and Abstract
Book.
Gaithersburg, Maryland, U.S.A.-P.80.
\bibitem{derzh2} Derzhko O.V., Moina A.Ph., Levitskii R.R. Bose commutation
rules approximation in spin systems theory//The Eighth International
Meeting on Ferroelectricity, 8-13 August 1993. Program Summary and Abstract
Book.
Gaithersburg, Maryland, U.S.A.-P.80.
\bibitem{Stein} Steiner M., Villain J., Windsor C.G. Theoretical and
experimental
studies on one-dimensional magnetic systems//Advances in Physics.-1976.-
{\bf 25}, 2.-P.87-209.
\bibitem{Mikeska} Mikeska H.-J., Steiner M. Solitary excitations in
one-dimensional
 magnets //Advances in Physics.-1991.-{\bf 40},3.-P.191-356.
\bibitem{Land} Landee C.P., Willett R.D. Tetramethylammonium copper chloride
and {\em tris} (trimethylammonium) copper chloride:
\sod\ Heisenberg one-dimensional ferromagnets//Phys.Rev.Lett.-1979.-{\bf 43},
6.-P.463-466.
\bibitem{Willett} Willett R.D., Landee C.P., Gaura R.M., Swank D.D.,
Groenedijk H.A., van Duyneveldt A.J. Magnetic properties of one-dimensional
spin \od\ ferromagnets: metamagnetic behavior of $(C_6H_{11}NH_3)CuCl_3$//
Journal of Magnetism and Magnetic Materials.-1980.-{\bf 15-18}.-P.1055-1056.
\bibitem{Tak} Takahashi M., Turek Ph., Nakazawa Y., Tamura M., Nozawa K.,
Shiomi D., Ishikawa M., Kinoshita M. Discovery of a quasi-1D organic
ferromagnet,
{\em p}-NPNN//Phys.Rev.Lett.-1991.-{\bf  67},6.-P.746-748.
\bibitem{Silv} Silverstein A.J., Soos Z.G. Comparison of three self-consistent
ground states for the linear Heisenberg antiferromagnet//J.Chem.Phys.-1970.-
{\bf 53},1.-P.326-333.
\bibitem{Car} De Carvalho A.V., Salinas S.R. Theory of phase transition
in the quasi-one-dimensional hydrogen-bonded ferroelectric crystal $PbHPO_4$//
J.Phys.Soc.Japan.-1978.-{\bf 44},1.-P.238-243.
\bibitem{Zin} Zinenko V.I. Theory of ferroelectrics of $ PbHPO_4$ type
//Fiz.Tv.Tela.-1979.-{\bf 21},6.-P.1819-1825. (in Russian)
\bibitem{Levi} Levitsky R.R., Grigas J., Zachek I.R., Mits Ye.V., Paprotny W.
Relaxational dynamics of quasi-one-dimensional $CsD_2PO_4$-type
ferroelectrics//Ferroelectrics.-1986.-{\bf 67}.-P.109-124.
\bibitem{Iorio1} D'Iorio M., Armstrong R.L., Taylor D.R. Longitudinal and
transverse
spin dynamics of a one-dimensional XY system studied by chlorine nuclear
relaxation in $PrCl_3$//Phys.Rev.B.-1983.-{\bf 27},3.-P.1664-1673.
\bibitem{Iorio1} D'Iorio M., Glaus U., Stoll E. Transverse spin dynamics of a
one-dimensional XY system: a fit to spin-spin relaxation data//Solid  State
Communications.-1983.-{\bf 47},5.-P.313-315.
\end{thebibliography}
\end{document}